\documentclass{article}
\pdfoutput=1

\usepackage{longtable}
\usepackage{graphicx}

\begin{document}                                                                                   
\centerline{\bf{Evolutionary studies of the Young Star Clusters: \\NGC~1960, NGC~2453 and NGC~2384 }}
\centerline{\bf{Priya Hasan, G. C. Kilambi and S. N. Hasan}} 
\centerline{\bf{Department of Astronomy, Osmania University, Hyderabad~-~500007,~India.}}

\date{\today}

\begin{abstract}
We report the analysis of the young star clusters NGC~1960, NGC~2453 and NGC~2384  observed in the $J$~(1.12~$\mu$m), $H$~(1.65~$\mu$m) and $K^{'}$~(2.2~$\mu$m) bands. Estimates of reddening, distance and age as $E(B-V)=0.25$, $d= 1380$ pc and $t=31.6$ to $125$ Myr for NGC~1960, $E(B-V)=0.47$, $d=3311$~pc and $t=40$ to 200 Myr for NGC~2453 and $E(B-V)=0.25$, $d=3162$~pc and $t=55$ to 125 Myr for NGC~2384 have been obtained. Also, we have extended the color--magnitude diagrams of these clusters to the fainter end and thus extended the luminosity functions to fainter magnitudes.  
The evolution of the main sequence and luminosity functions of these clusters have been compared with themselves as well as Lyng\aa \  2 and NGC~1582.        
\end{abstract}
{\bf{keywords: star clusters:young -- near-infrared photometry -- 
       color--magnitude diagrams -- pre-mainsequence stars --
      2MASS}}

\section{Introduction}

 In recent times, the use of infrared array detectors has led to a major revolution in the study of young clusters and star forming regions. Star clusters are the sites of star formation where gravitationally bound systems of stars are formed from the same parent cloud. They are classic research objects in a wide range of stellar and galactic investigations. Color--magnitude diagrams of galactic clusters have been the backbone of studies of stellar formation and evolution as clusters provide a ready sample of stars at the same distance, of the same chemical composition and age, differing only in mass (Friel 1995).

With these objectives in mind, an extensive study of young star clusters was initiated using the Mt Abu Infrared Observatory, Gurushikhar, Rajasthan, India. In this work, we present the observations and analysis of the young star clusters NGC~1960, NGC~2453 and NGC~2384 observed in February 2000 in the near infrared $J$~(1.12~$\mu$m), $H$~(1.65~$\mu$m) and $K^{'}$~(2.2~$\mu$m) bands using the NICMOS3 detector at the Cassegrain focus of the 1.2~m telescope.   Preliminary results for these clusters have been presented (Hasan, Kilambi \& Baliyan 2001, 2002, 2003 and Hasan 2005a, 2005b). This article offers a complete and comprehensive analysis of these clusters and compares the stellar population of these clusters with clusters of comparable age {\it viz.} Lyng\aa~\ 2 and NGC~1582.
 

 Basic parameters of these clusters like distance, reddening and age have been estimated using optical data earlier. The Two Micron All Sky Survey (2MASS) data is available for these clusters in the $JHK$ passbands. But, so far it has not been used to obtain cluster parameters.  These would be the first results obtained using $JHK$ data on these clusters. The parameters of these clusters catalogued in Lyng\aa \  (1987) are given in  Table~\ref{clusterdata}.

\begin{table}[h]
\small
\begin{tabular}{llll}
\hline
Parameter & NGC~1960 & NGC~2453 & NGC~2384 \\ \hline 

RA(2000)(h:m:s) &5 36 6 & 7 47 45 & 7 25 13\\
Decl.(2000)(d:m) & 34 08& -27 14 &-21 01 \\
Galactic longitude(deg) &174.52 & 243.33  &235.39 \\
Galactic latitude(deg) &+1.04 &-0.94 &-2.41 \\
Trumpler class & &I 3 m &IV 3 p \\
Ang.diameter(arc min) &10.0 &4.0 &5.0 \\
Distance(pc) &1270 &1500 &2000 \\
$E(B-V)$(mag) &0.24 &0.47 &0.29 \\
log(age) &7.4 &7.6 &6.00 \\
Radial velocity(km s$^{-1}$) &-4 & &67.3 \\
Linear diameter(pc) &2.2 &1.1 &1.5 \\ \hline
\end{tabular}
\caption{Basic cluster parameters Lyng\aa \  (1987)}
\label{clusterdata}
\end{table}

 NGC~1960 (M 36) has been studied by Boden (1951), Johnson and Morgan (1953) and Barkhatova {\it et al.}~(1985).
CCD photometry has been published by Sanner {\it et al.} (2000). 
The proper motions of stars in the cluster have been presented by Meurers (1958), Chian \& Zhu (1966) and Sanner {\it et al.}~(2000).  Radial density profiles of this and 37 other open clusters in the core and corona have been obtained by Nilakshi {\it et al.}~(2002).

NGC~2453 is located in Puppis and is  moderately close to the clusters Haffner 18 and 19 which have been identified with a possible spiral arm at 15 kpc from the galactic center by Moffat \& Fitzgerald (1974). 
The cluster has been studied by Seggewiss (1971), Moffat \& Fitzgerald (1974) and Fitzgerald, Luiken \& Maitzen (1979) with varied distance estimates. 
Gathier (1984)  and  Mallik, Sagar  \& Pati (1995) have also studied this cluster, as it lies in close proximity of the planetary nebula NGC~2452, to determine an improved reddening--distance relation for the nebula and to settle unambiguously the question of the distance to the cluster and that of its physical connection to NGC~2452. 

NGC~2384 has been studied by Trumpler (1930), Collinder (1931), Vogt and Moffat (1972), Hassan (1984),  Babu (1985) and Subramaniam and Sagar (1999).

Both the clusters NGC~2384 and NGC~2453 lie in the third galactic quadrant in the Canis Major--Puppis--Vela region. Moitinho (2001) has obtained $UBVRI$ photometry of these and other 28  open clusters to study the star formation history and spatial structure of this region. 

\section{Observations and Data Reduction}
   
  Observations were made on 5 -- 7 February 2000 with the NICMOS3 camera which is a HgCdTe $256 \times 256$  pixel
array (Joshi {\it et al. }~2002).
The typical exposure times for the $J$, $H$ and $K^{'}$ bands were typically 30, 10 and 3 seconds respectively and the typical number of exposures were 5, 11 and 21 respectively. 
A complete description of the NICMOS3 and photometry procedures using the same have been presented in Baliyan {\it et al.}~(2002). 
  
  The field of view $2'\times 2'$  was used in these observations and the telescope was 
moved in raster form to scan the entire  cluster area at 25-30 locations. The average seeing was between $1.5''$ to $2''$. Photometry was performed on individual tiles as the background varied very strongly. The stars in overlapping locations were cross-identified manually to obtain a final photometry file and the images were later mosaiced to obtain the total cluster area. From the resulting files of photometry, we have deleted all objects showing high photometric errors, sharpness and $\chi$ values for which the limits were chosen individually for each image (typical values were $0^{m}.03$ to $0^{m}.05$ for the magnitudes, $\pm$0.5 to 1 for sharpness and 2 to 4 for $\chi$). Cleaning and photometry of the images was done using the Image Reduction and Analysis Facility (IRAF)  and its external package DAOPHOT (Peter Stetson 1987, 1992) made available to the astronomical community by the National Optical Astronomy Observatories (NOAO), USA. 

Standard stars from Hunt {\it et al.}~(1998) were observed. The standards were observed at different locations in the detector but very close to the zenith and the average magnitudes were used. The transformation equations obtained using the data are as follows:
$$J=J_{i}+0.0242(\pm 0.0053)(J-K^{'})-5.4319(\pm 0.0229)$$
$$H=H_{i}+0.1279(\pm 0.0143)(J-H)-5.6591(\pm 0.0528)$$
$$K^{'}=K_{i}-0.2683(\pm 0.068)(J-K^{'})-6.5126(\pm 0.0325)$$
where $J$, $H$ and $K^{'}$ are the magnitudes in the standard system, $J_{i}$, $H_{i}$ and $K_{i}$ are the magnitudes in the instrumental system. These relations show a small color dependence for the  $J$ magnitude, but a substantial color dependence in the case of $H$ and $K^{'}$ magnitudes (especially for late type stars). 

The observed data for all the stars in the standard magnitude 
system for NGC~1960, NGC~2453 and NGC~2384 will be available in electronic form through CDS, Strasbourg or on request from the author.




Figures ~\ref{alljjh1},~\ref{alljjh2} and ~\ref{alljjh3} show the apparent color--magnitude diagrams $J$, $H$ and $K^{'}$ versus $(J-H)$ respectively, for all stars 
observed in the fields of the clusters, where  $J$, $H$ and $K^{'}$ are the observed magnitudes and $(J-H)$ is the observed color. 
 
\section{Cluster radius and membership}
\label{clusrad}
As the fields observed were small, the data from 2MASS has been used to study the extent of  the clusters. 2MASS has uniformly scanned the entire sky in the same photometric bands, i.e.  $JHK$ and hence a wide coverage of the cluster and its surrounding field can been  obtained (Skrutskie {\it et al.} 2006). The point-source $S/N$ =10 limit is acheived at or fainter than $J=15^{m}.8, H=15^{m}.1 $ and $K_{s} =14^{m}.3$  for virtually the entire sky. Hence, we constrained the data used by these brightness limits and only used objects showing photometric errors $\leq 0^{m}.2$.

 The centers of the clusters are determined iteratively by calculating the average $X$ and $Y$ positions of the 
stars within 300 pixels from an eye estimated center, until they converged to a 
constant value. An error of a few tens of pixels or a few arc seconds (plate scale for 2MASS is $2'' $pixel$^{-1}$) is expected in locating the center.  Following Kaluzny (1992), attempts were made to fit 
the plot to $$\rho(r)=\frac{f_{0}}{1+(r/r_{c})^2}$$ where $r_{c}$ the cluster's core radius  
is the radial distance at which the value of $\rho(r)$ becomes half of the 
central density $f_{0}$.  Using the $\chi ^{2}$ minimization technique  $r_{c}$ and other constants were determined. The radial density plots for the clusters(solid lines) with the corresponding fits(dotted lines) are shown in Fig.~\ref{allrad}.
The limiting radius of the cluster is the distance from the centre at which the star density becomes approximately equal to the field star density. Table \ref{radall} lists the coordinates of the cluster centers for epoch 2000, limiting radius obtained, radius as in Dias, Alessi,  Moitinho \&  Lepine (2002) and fraction of the cluster observed. 

\begin{table}[h]
\small  
\begin{tabular}{ccccc}
\hline
Cluster & Center Coords &  limit radius & Dias {\it et al.}(2002) & Fraction   \\ 
 &  $\alpha_{2000}$, $\delta_{2000}$ & arc min & arc min & observed\\ \hline
NGC~1960 & $5^{h}36^{m}12^{s}$  $+34^{0}8^{m}6^{s}$ & $10'$  & $10'$ &0.7 \\
NGC~2453 &$7^{h}47^{m}32^{s}$  $-27^{0}14^{m}25^{s}$ & $5'$ &$4'$  &0.95 \\
NGC~2384 & $7^{h}24^{m}59^{s}$  $-21^{0}1^{m}6.6^{s}$ & $5'$  & $5'$ &0.85 \\ \hline
\end{tabular}
\caption{Cluster centers and sizes}
\label{radall}
\end{table}


The cluster membership has been established using proper motion data, if available, and also from the radial distance technique and photometric criterion. The photometric method described by Walker (1965) involves plotting 
all the observations in the $m_{J_{0}}$ -- $M_{J}$ plane where $m_{J_{0}}$ is the apparent unreddened magnitude and $M_{J}$  is the absolute magnitude  (Fig.~\ref{evol1}). A straight line 
representing the adopted distance modulus is drawn. In this diagram, it is assumed that an unresolved binary could cause a maximum deviation from the line by $0^{m}.75$ with equal components. It is further assumed that the observational scatter could cause a vertical displacement of not more than $0^{m}.5$ for stars appearing on the main sequence. These two boundaries are represented by dashed lines in Fig.~\ref{evol1}. All stars lying within these boundaries and also on the border areas are treated as  members. This method will be further refered to as the evolutionary track method.
 A small error in estimation of the distance modulus will not lead to misidentification of a large number of members.  The evolutionary track method identifies only main sequence stars while other luminosity classes and  groups require other methods for member identification.

For NGC~1960, cluster membership was established using the proper motion data of Sanner {\it et al.} (2000) (47 stars) and also through the evolutionary track diagram  or the photometric technique (187 stars), thus increasing membership to 234 stars. 
In addition, 3 more stars, suspected to be $T$ $Tauri$ stars, identified using a technique described in the section on the two-color diagrams have also been added. 


There is substantial field star contamination in the region of NGC~2453 and in the absence of kinematical data, establishing cluster membership is a tedious task. Cluster membership was established based on the basis of the  probable and confirmed members (spectroscopically) identified by Moffat \& Fitzgerald (1974) (30 stars) and also from the evolutionary track method (108 stars).

Cluster membership for NGC~2384 was established using the members identified by Vogt and Moffat (1972) (8 stars)  and also through the evolutionary track method (111 stars).


\section{Reddening Correction and Distance Modulus}

In our analysis, the estimate of the inter and intra-cluster  extinction were carried out using the available spectroscopic and photometric data in the optical and $JHK$ bands. 

The observed data was then corrected for interstellar reddening using the 
coefficients given by Bessell and Brett (1988) where $A_{J} = 0.96 \times E(B-V), A_{H} = 0.80 \times E(B-V), A_{K} = 0.12 \times E(B-V), E(J-H) = 0.37 \times E(B-V), E(H-K) = 0.19 \times E(B-V)$ and $E(B-V)$ is the color excess adopted for the cluster.

\subsection{NGC~1960}
For NGC~1960, the spectroscopic data of Johnson and Morgan (1953), Hiltner (1956), Abt and Morrell (1995) and Chargeishivili (1988) has been used.
 
 The reddening determined by isochrone fitting to the color--magnitude diagram is $E(B-V)=0.25$ which agrees with the value of $0.20$ obtained by Cummings {\it et al.}~(2002) and with that of $0.25$ obtained by Sanner {\it et al.} (2000).

The spectroscopic data available in literature was used to obtain an estimate of 
the distance modulus of this cluster. 
 Using the spectral classification of Johnson and Morgan (1953), 
we get an apparent distance modulus of 11.43 $\pm$0.52 (true distance modulus 
10.37 $\pm$0.52). From the differential reddening plot, the true distance modulus was estimated as 10.58 $\pm 0.18$.

For the present analysis, the distance to the cluster was obtained through 
sliding fit of the standard main sequence with that of the color--magnitude 
diagram and it gave  a true distance modulus of $10^{m}.7$ corresponding to a 
distance of 1380 pc.

\subsection{NGC~2453}
The only spectroscopic data available for NGC~2453 is for MF 54, which is a confirmed member with a spectral class of $B5 V$ (Moffat \& Fitzgerald 1974). According to the authors, this star is an evolved blue giant and is also a spectroscopic binary with a radial velocity of 67 $\pm 14$ k s$^{-1}$. Hence any conclusions of reddening and distance modulus cannot be made based on the spectroscopy of this object.

By the method of isochrone fitting, we have obtained a color excess of 0.47 which agrees with that obtained by Moffat \& Fitzgerald (1974) and Mallik {\it et al.} (1995).
According to the extinction--distance diagrams presented by Neckel and Klare (1980), the galactic region of the cluster has nearly a constant $A_{V}$ of $1^{m}$ to about 5 kpc and then an increase to $2^{m}$  at 6 kpc and beyond and it agrees with our estimates. 
For a more reliable result, we have used the main sequence fitting technique, and obtained a true distance modulus of 12$^{m}$.33 corresponding to a distance of 3311 pc.

Mallik {\it et al.} (1995) suspected a possible physical connection of the cluster NGC~2453 with the planetary nebula NGC~2452 for the following reasons:

1. The radial velocity of NGC~2452 was measured by Campbell \& Moore (1918) to be 68~km~s$^{-1}$ and the radial velocity of an evolved blue giant MF 54 in the cluster field was estimated as 67~km~s$^{-1}$.

2. The $E(B-V)$ of the cluster and nebula are 0.47 and 0.43 respectively.

Inspite of almost similar values of radial velocity and extinction, Mallik {\it et al.} (1995) ruled out this possibility as they obtained a distance of 5.9 kpc to the cluster, considering it to be a mere coincidence. But, as we have obtained a much smaller distance, the likelihood of the association of the cluster with the nebula whose largest distance to date is 3.47~kpc (Gathier 1984), needs to be reconsidered.

\subsection{NGC~2384}

In the case of NGC~2384, the spectroscopic data of Walker (1956), Garrison (1977), Subramaniam and Sagar (1999), Babu (1985),  Fitzgerald {\it et al.}~(1979) and Houk, Cowley and Smith-Moore (1975) has been used.

Using the isochrone fitting method, we found a reddening value of 0.25, which agrees well with that of Subramaniam and Sagar (1999).
Between galactic latitudes 231$^{0}$ $\leq $ l $\leq$ 256$^{0}$, the interstellar absorption is small up to large distances. The reddening increases slowly upto a distance of 1 kpc, and beyond that it is nearly a constant $E(B-V)=0.3$ upto more than 4 kpc (Isserstedt \& Schmidt-Kaler 1964,  Neckel \& Klare 1980). This also supports the value obtained.

For the stars which were identified members by Vogt and Moffat~(1972) and for which spectroscopic data is available, the mean true distance modulus has been obtained as $12.^{m}6$ in $V$ band (using the data of Vogt and Moffat 1972) and $11.^{m}96$ in $J$ band (using our data) 
Using  the sliding fit method, we obtained a true distance modulus of $12^{m}.5 \pm 0.5$ corresponding to a distance of 3162~pc. 

\section{Discussion}

\subsection{NGC~1960}

The unreddened color--magnitude diagrams for all members are shown in Fig. 
\ref{iso1}. All proper motion stars are 
denoted by filled circles (47 stars), evolutionary track members by open circles 
(187 stars) while possible $T$ $Tauri$ candidates by filled triangles (3 stars). The dashed line represents the standard main sequence (Koornneef 1983) for normal stars.
The color--magnitude diagram of the cluster has been further extended to  
$M_J \approx 5^{m}$ through the evolutionary track technique. The 
distribution of stars range from spectral types late $O$ or early $B$ types to 
$M$ types or even later. There seems to be a binary main sequence formed by 
unresolved binaries above the main sequence, as already suggested by Sanner {\it et al.} (2000). Not all 
stars lie above the main sequence as observed by Sanner {\it et al.} (2000), 
probably due to difference in phases of the observed binary at the time of the 
two observations or due to unequal components of the binaries. One of the stars 
(Id No 190), suspected by Sanner {\it et al.} (2000) as a member, was not 
identified by us as a member using our techniques. 
The main sequence is narrow for early type stars and it gradually broadens for late type stars. 
Slettebak (1985) has reported
two $Be$ stars. One of them is Boden number 101, observed by Johnson and Morgan (1953) which is probably a member (Sanner {\it et al.} 2000). The second
is Boden number 505 (erroroneously named 504 by Slettebak) from photometry by
Hiltner (1956) lies near the edge of the cluster. Though
its position in the color--magnitude diagram is consistent with membership, yet
it is suspected to be a non-member as the proper motion data fails to satisfy
membership criterion.  Both these stars do not lie in our observed field and
hence we are unable to make any conclusions about them. The star Id. No 473 
(Sanner~No~1) is the brightest star in the cluster and almost falls on the standard main
sequence. As it is the most massive star, its actual 
position should be to the right of the main sequence due to evolution. But, its 
observed position could be due to free-free emission from the electrons in 
its shell which cause the star to move towards the left in the HR diagram, or due to its possible variable nature. Further observations are necessary to ascertain the real cause of its observed position in the HR diagram. 


It has been found that $T$ $Tauri$ stars, the stars which are still in the 
gravitational contraction stage, occupy a certain narrow region in the 
$(J-H)_0$--$(H-K)_0$ plane (Rydgren \& Verba 1983).  A parallelogram
 was constructed to describe this region, using the data from the 
above reference. 
The color--color diagram $(J-H)_0$ versus $(H-K)_0$ for NGC~1960  members is 
plotted in Fig.~\ref{red1}. The standard main sequence is shown as a solid curve 
plotted from Bessell and Brett (1988). On taking into account the observational 
scatter, the distribution of some late type stars suggest that they have very 
large infrared excesses (as seen from the plot) due to the young nature of the 
cluster. In Fig.~\ref{red1} the proper motion members are denoted by filled circles and 
evolutionary track members by open circles. Also indicated in the diagram is the 
narrow track occupied by $T$ $Tauri$ stars. Stars which lie in the $T$ $Tauri$ 
parallelogram  can be considered to be probable members of $T$ $Tauri$ type 
from the observed colors (Id Nos 384, 399 and 411), depending upon their position in the color--magnitude 
diagram and their position from the center of the cluster.

Young clusters seem to have a spread in age and hence more than one 
isochrone would fit the data. The brightest star (No 473), as is noticeble in the plot, clearly lies far away from the majority of the other bright stars in the cluster. An isochrone of 31.6 Myr (Bertelli {\it et al.}~(1994), metallicity =  0.02) would fit its position (Fig.~\ref{iso1}).
The observed position of star  No 473  is puzzling from the evolutionary point of view. Based on the position of the remaining bright stars one could estimate an age of 100--150 Myr (mean age 125 Myr) rather than accept a younger age on the basis of the position of a single star (Fig.~\ref{iso1}). 

In Fig.~\ref{iso1}, we have compared the color-magnitude diagram of NGC~1960 with that of Lyng\aa~\ 2 (indicated by squares) which has a true distance modulus of $9.7 \pm 0.1$, E(B-V)=0.22 and an age of 90 Myr (Bica{\it et al.}~2006). A close examination of the color-magnitude diagrams suggest that both have a similar structure and also more or less the same evolutionary status  excluding star No 473  of NGC~1960. 

\subsection{NGC~2453}

Figure \ref{iso2} shows the unreddened color--magnitude diagram for NGC~2453 in $JHK$ passbands. The stars which Moffat \& Fitzgerald (1974) identified as members are marked with filled circles while the members by evolutionary track method are marked with open circles. From the plot, it is noticeable that the cluster is not very young and stars of absolute magnitudes $J \leq  -0.5^{m}$ seem to have already begun moving away from the main sequence. There are three stars of particular interest as indicated by Moffat \& Fitzgerald (1974). The first is MF~54 (Id No 410), a blue giant showing double spectral lines (as discussed earlier), the second is MF~40 (Id No 376) which shows strong $H_{\alpha}$ emission and  the third MF 50 (Id No 513)  is a bright red giant and a probable member. MF 50 has been identified as a variable star and a member by Lapasset, Claria and  Minniti (1991) and lies away from the main sequence. MF~54 and MF~40 have also been identified as  Be type variable stars.



 The isochrones from Bertelli {\it et al.} (1994) for metallicity 0.02 have been used as in the case of the previous cluster. To determine the age, if the star No 513 is used, it leads to an age of 40 Myr (Fig. \ref{iso2}).

The estimate of age of 40 Myr or less by Mallik et al (1995) based on the average of the isochrone fits to MF 50 and MF 40 seems to be somewhat puzzling as majority of the brighter members fit to larger age isochrones. If the age is determined using the star number 410 (MF 54), an age given by 158 Myr is obtained. But both MF 50 and MF 54  are variables and hence neither of the two ages can be reliable. The isochrone of 251 Myr fits the next set of brighter stars. We can thus state that the age of the cluster lies between 158 -- 250 Myr with the mean age as 200 Myr. 

In Fig. \ref{iso2}, we have overlayed the color-magnitude diagram of NGC~1582 (indicated by squares) using 2MASS data (Baume, Villanova  \& Carraro 2003). This cluster is at a distance of 1100 pc, E(B-V) = 0.35 and has an age 200 $\pm$ 100 Myr. There seem to be more bright stars in NGC~2453 compared to NGC~1582. In NGC~1582, there seems to be a  paucity of stars between $2^{m} \leq J \leq 3^{m}$, unlike NGC~2453.

\subsection{NGC~2384}

The color--magnitude diagram for NGC~2384 is shown in Figure \ref{iso3}. Evolutionary track members are shown with open circles, while filled circles show members identified by Vogt and Moffat (1972).  .

From the color--magnitude diagram, we see that the main sequence is fairly broadened which may be due to the probable presence of a number of binaries. Some of the stars identified by Vogt and Moffat (1972) are spectroscopic binary members and hence do not sit on the main sequence. 


Determination of the age of NGC~2384 is difficult as there are not too many advanced evolved stars in the cluster. If  we were to determine age on the basis of the brightest star, the age would be around 50 Myr. Based on the majority of the bright stars' position, there appears to be an age spread from $\approx$ 55 Myr to 125 Myr, with an average of 90 Myr.  Figure \ref{iso3} shows the isochrones fit to the color--magnitude diagram of the cluster as well as the plot for Lyng\aa~\ 2 (indicated by squares). The lower mass members of NGC~2384 seem to be above the main sequence implying that they are still in the gravitational contraction phase, unlike in Lyng\aa~\ 2 . There are a few more massive young stars present in NGC~2384 evolving from  the main sequence while it is not the case with Lynga 2, suggesting a somewhat younger age for NGC~2384. Subramaniam and Sagar (1999) found a much younger age for this cluster as they have not considered the stars identified by Vogt and Moffat (1972) as members based on spectroscopy.

Table \ref{allpar} shows the parameters obtained by us compared to those determined by previous authors for all the three clusters. 

\begin{table}[h]
\small
\begin{tabular}{llll}
\hline
   Cluster &  Reference  &   Distance(kpc) & Age(Myr)  \\
             \hline
  
  NGC~1960 &  Barkhatova {\it et al.}~(1985)  &       &31.6 \\
           &  Boden (1951)   & 1.1 \\
           & Sanner {\it et al.} (2000) & 1.3   &16\\
           & This work       &  1.13 & 125\\ 
                       \hline
			       
 NGC~2453          &Moffat \& Fitzgerald (1974) &  2.9 &  40 \\
           &Gathier (1984)              &  5.0 &  \\
           &Mallik {\it et al.} (1995)         &  5.9  &   25\\
           &This work                   &  3.3  &  200\\
\hline

NGC~2384   &Vogt and Moffat (1972)      &  3.28 &         \\
           &Subramaniam and Sagar (1999)&  2.93 & 15.8 \\
           &This work                   & 3.16  & 90\\  \hline 
			     	       
\end{tabular}
\caption[]{Parameters estimated for NGC~1960, 2453 and 2384}
\label{allpar}
\end{table}

As seen in the HR diagram of these clusters, there are suspected gaps corresponding to the Memilliod gap (Memilliod 1976)  at $B8V$ ($(J-H)_{0} = -0.05$), the Canterna gap (Canterna, Percy \& Crawford 1979) at $(J-H)_{0} = 0.01$ and the M11 gap at  $(J-H)_{0} = 0.04$ for early $A$ types. These are possibly related to the way the Balmer jump and Balmer lines behave in late $B$ and early $A$ stars. There is also a possible Bohm--Vitense gap (Bohm--Vitense and Canterna 1974) at $(J-H)_0=0.13$ and also a possible gap at $(J-H)_0=0.32$. 
Though the apparent distribution of stars on the main sequence suggests the presence of gaps, the real identity of these gaps requires a detailed analysis. We have used the method of Hawarden (1971) and found them to be statistically insignificant. This needs to be further investigated since gaps are very good indicators of possible physical processes in stars. 
To determine the true luminosity function of a cluster, it is necessary to 
correct the observed data for completeness, field star contamination and the 
fraction of the observed cluster.


Figure \ref{alllf} shows the uncorrected (dotted line) and corrected (solid line) 
luminosity functions for the three clusters. In the case of NGC~1960, the absolute magnitude 
 ranges from $-2^{m}.5$ to $5^{m}.5$ with a peak near $4.0^{m}$. 
Though it is a fairly young cluster, the luminosity function suggests there are 
a few bright stars and many more intermediate massive stars. 

NGC~2453 is a relatively older cluster as it has some evolved members. The majority of stars range from magnitudes $-2^{m}.5$ to $3^{m}.5$, while the peak lies somewhere near $1^{m}$. The absolute magnitudes of the members of NGC~2384 range from $-3^{m}.5$ to $4^{m}.5$, with a peak at around $2^{m}$. 
The sudden drop at fainter magnitudes is due to the observational limitations at present.


Figure \ref{2m1960comp} shows the comparison between the magnitudes and colors obtained by 
2MASS and our observations for NGC~1960. . 
It is noticeable in the plot that there appears to be an offset  
 which could be due to errors in the photometric 
calibration of the 2MASS data, mismatch of filters or atmospheric conditions. 
But, as this offset varies from cluster to cluster, we assume that this is due to 
atmospheric conditions. Transformations to many present systems have been studied by Carpenter (2001). 
 \section{Conclusions}

The clusters NGC~1960, NGC~2453 and NGC~2384 have been studied in the $JHK$ passbands to  obtain their parameters which are listed  in Table \ref{allpar}.

NGC~1960 and NGC~2384 have similar ages (125 and 90 Myr) and sizes (4 pc and 4.6 pc respectively).
The cluster NGC~2453 has a linear size is 3.85 pc though it is much older (200 Myr). 
A very detailed study was made of the color-magnitude diagrams of all these clusters and compared with Lyng\aa~\  2 and NGC~1582. A comparative study of the luminosity functions  of all the observed clusters has shown that the luminosity function peaks at $2^{m}$ for NGC~2384, $4^{m}$ for NGC~1960 and at $1^{m}$ for NGC~2453. 

We need to study a yet larger sample of young clusters to see how these characteristics vary with age, position in the galaxy and gas content.

  \section{Acknowledgements}
     The valuable comments of the referee have helped in greatly improving the scientific content of this paper. The authors are grateful to K S Baliyan and the Physical Research Laboratory, India for providing valuable telescope time. PH is also grateful to Inter University Centre for Astronomy and Astrophysics, Pune for providing library and computing facilities. This publication makes use of data products
from the Two Micron All Sky Survey. This work has also made use of the WEBDA database, operated at the Institute for Astronomy at the University of Vienna (http://www.univie.ac.at/webda) founded by J.-C.Memilliod devoted to observational data on galactic open clusters. This research has been funded by the Department of Science and Technology (DST), India under the Women Scientist Scheme (PH).


\section{Bibliography}

Abt H.A., Morrell N.I., \newblock{ApJS}, {\bf 99}, 135, 1995.\\
 
Babu G.S.D.,  \newblock{PhD Thesis, Bangalore University, India}, 1985.\\

Baume G., Villanova S. and Carraro G. \newblock{Astron. Astrophys}, {\bf 407}, 527, 2003.\\

Baliyan K.S., Sanchawala K., Ganesh S., Joshi U.C., Shah C.R.,  
in: Automated Data Analysis in Astronomy,eds. Ranjan Gupta, H.P.Singh and C.A.L Bailer--Jones, Narosa Publishing House, India, 2002.\\
 
Barkhatova K.A., Zakharova P.E., Shashkina L.P., Orekhova L.K., \newblock{PAZh}, {\bf 62}, 854, 1985.\\


Bertelli G., Bressan A., Chiosi C., Fagotto F., Nasi E.,  
\newblock{A\&AS}, {\bf 106}, 275, 1994.\\

Bessell M.S., Brett J.M.,\newblock{PASP}, {\bf 100}, 1134, 1998.\\

Bica E.,  Bonatto C., Blumberg R., \newblock{A \& A}, {\bf 460}, 83.\\

Boden E.,  \newblock{Uppsala Astr. Obs. Ann. 3}, No 4, 1951.\\

Bohm--Vitense E., Canterna R., \newblock{ApJ}, {\bf 194}, 629, 1974.\\

Campbell, \newblock{Publ Lick Obs}, {\bf 13}, 75, 1918.\\

Canterna R., Percy C.L., Crawford D.L., \newblock{PASP}, {\bf 91}, 263, 1979.\\

Carpenter B.S., \newblock{AJ }, {\bf 121}, 2851, 2001.\\

Chargeishivili K.B., \newblock{Abastumai Astr. Obs. Bull.}, {\bf  65}, 1, 1988.\\

Chian B., Zhu G., \newblock{Ann. Shesan G. Sect. of Shanghai Obs.}, 26, 53, 1966.\\

Collinder P., \newblock{Ann Lund Obs}, {\bf 2}, 1931.\\

Cummings J., Jacobson H., Deliyannis C.P., Stenhauer A., Sarajedini A., \newblock{Bulletin of the American Astronomical Society }, {\bf 34}, 1308, 2002.\\

Dias W.S., Alessi B.S., Moitinho A., Lepine J.R.D., \newblock{A\&A}, {\bf 389}, 871, 2002.\\


FitzGerald M.~P., Luiken M.,   Maitzen H.~M. and  Moffat A.~F.~J., \newblock{A \& AS}, {\bf 37}, 354, 1979.\\

Friel E. D.\newblock{Ann. Rev. Astron. \& Astroph.}, {\bf 33}, 381, 1995.\\

Garrison R. F.,   Hiltner W. A. and Schild R. E., \newblock{Ap J. S. S.}, {\bf 35}, 111, 1977.\\

Gathier R.,  \newblock{Ph D thesis, University of Groningen}, 1984.\\


Hasan P., Kilambi G. \& Baliyan K S., \newblock{BASI}, {\bf  29}, 329, 2001.\\

Hasan P., Kilambi G. \& Baliyan K S., \newblock{BASI}, {\bf  30}, 653, 2002.\\

Hasan P., Kilambi G. \& Baliyan K S., \newblock{BASI}, {\bf  31}, 383, 2003. \\

Hasan P.,  \newblock{BASI}, {\bf  33}, 151, 2005a.\\ 

Hasan P., \newblock{BASI}, {\bf 33}, 311, 2005b.\\

Hassan S. M., \newblock{Astronomy with Schmidt-Type Telescope}, Dordrecht: Reidel, {\bf 295}, 1984.\\


Hawarden., \newblock{Observatory}, {\bf 91}, 78, 1971.\\

Hiltner W.A., \newblock{ApJS}, {\bf 2}, 389, 1956. \\

 
Houk N. ,  Cowley A.P. and  M. Smith-Moore, \newblock{IAU Coll}, {\bf 32}, 357, 1975.\\


Hunt L.K., Manucci F,  Testi L.,  Migliorini S.,  Stanga R. M.,
Baffa C.,  Lisi F., Vanzi L., \newblock{AJ }, {\bf 115}, 2594, 1998.\\

Isserstedt J. and Schmidt-Kaler T., \newblock{Z. Astrophys.}, {\bf 59}, 182, 1964.\\

Johnson H.L., Morgan W.W.,  \newblock{ApJ}, {\bf 117}, 313, 1953.\\

Joshi U.C., S Ganesh, Baliyan K.S., Shah A.B., Vadher N.M., 
Automated Data Analysis in Astronomy Ed. Ranjan Gupta, H.P.Singh and C.A.L Bailer--Jones, Narosa Publishing House, India, 2002.\\

Kaluzny J., Udalski A., 1992, \newblock{Acta Astron.}, {\bf 42}, 29, 1992.\\


Koornneef J., \newblock{A\&A}, {\bf 128}, 84, 1983.\\

Lapasset  E.,   Claria J.J. and  Minniti D. , \newblock{Information Bulletin on Variable Stars,Commission of the IAU}, {\bf 3594}, 1991.\\



Lynga \ G., \newblock{Catalogue of Open Cluster Data, 5th edn., 1987.\\ 

Mallik D.C.V.,  Sagar R. and  Pati A., \newblock{Astron. Astrophys. Suppl. Ser.}, {\bf 114}, 537, 1995.\\

Memilliod J.-C., \newblock{A\&A}, {\bf 53}, 289, 1976.\\


Meurers J.,  \newblock{Z Astrophys},  {\bf 44}, 203, 1958.\\

Moitinho A., \newblock{A\&A}, {\bf 370}, 436, 2001.\\

Moffat A.~F.~J. and  FitzGerald M.~P., \newblock{Astron. Astrophys. Suppl. Ser.}, {\bf 18}, 19, 1974.\\


Neckel Th. and Klare G., \newblock{Astron. Astrophys. Suppl. Ser.} , {\bf 42}, 251, 1980.\\

Nilakshi, Sagar R., Pandey A.K., Mohan V., \newblock{A\&A }, {\bf 383}, 153, 2002.\\



Rydgren A.E., Verba F.J., 1983, \newblock{AJ 88}, 1017, 1983.\\ 

Sanner J., Altmann M., Bruzendork J., Geffert M., \newblock{A\&A}, {\bf 357}, 471-483, 2000.\\


Seggewiss  W., \newblock{Veroffentl. Astron. Inst. Bonn.}, {\bf 83},  21, 1971.\\
 
Skrutskie, M. et al, \newblock{Ap J}, {\bf 131}, 1183, 2006.\\

Sletteback A.,  \newblock{ApJS}, {\bf 59}, 769, 1985.\\

Stetson P.B.,  \newblock{PASP}, {\bf 99}, 191, 1987.\\

Stetson P.B., ASP Conf. Ser., Astronomical Data Analysis Software and Systems I, eds. Worrall D.M., Biemesderfer C. and Barnes J., {\bf 25}, 1992.\\

Subramaniam A. and  Sagar R., \newblock{AJ}, {\bf 117}, 937, 1999.\\

Trumpler R.J., \newblock{Lick Obs. Bull.}, {\bf 420}, 14, 154, 1930.\\

Vogt N. and Moffat A.~F.~J., \newblock{Astron. Astrophys. Suppl. Ser.}, {\bf 7}, 133, 1972.\\

Walker M.,\newblock {ApJS}, {\bf 2}, 365, 1956.\\

 Walker M.,  \newblock{ApJ}, {\bf 141}, 660, 1965.\\
}

\end{document}